# Acoustic Pressure Monopole, Velocity Monopole, and Hybrid Monopole

Suet To Tang, Xiaonan Zhang, Chong Meng, and Z. Yang[*]

Department of Physics, Hong Kong University of Science and Technology, Clear Water Bay, Kowloon, Hong Kong, China

**Abstract**

We show that there are two types of acoustic monopoles, namely the pressure monopoles (PM's) and the velocity monopoles (VM's). Under one wave incidence, only the pressure field near a PM at resonance possesses monopolar symmetry, while for a VM at resonance only the velocity field possesses monopolar symmetry. It is the dipolar response function of the monopoles that dictates whether they are PM's or VM's. A hybrid monopolar device with neither velocity nor pressure field symmetry is also demonstrated by combining a VM and a PM, which exhibits two monopole resonances with neither PM nor VM symmetry. The breaking of symmetry could lead to higher absorption of the subwavelength scale devices.

Symmetry is a fundamental property of acoustic metamaterials (AM's), which have attracted extensive attention over the past years [1, 2]. Acoustic coherent perfect absorption (CPA) and acoustic coherent perfect channeling (CPC) have been realized by utilizing the monopolar and dipolar symmetry of the scatterers [3, 4]. In CPA, on-and-off contrast ratio exceeding 900 in total absorption has been experimentally realized by adjusting the relative phase of the two incident waves between 0 and $\pi$. In CPC, a monopole scatterer completely channels the two incident waves in the main waveguide into the side waveguide when the symmetry of the incident waves matches that of the scatterer, with an on-and-off channeling intensity ratio exceeding $2.6 \times 10^4$ [4]. Symmetry could also lead to constraints of acoustic properties of AM's. It has been shown that structures with front-and-back mirror structural symmetry in Fig. 1(a-i) are either dipoles or monopoles [5, 6]. Slight breaking of such symmetry could lead to highly asymmetric metasheets with one surface totally absorbing and the other surface perfectly reflecting [7 – 9]. Topologically non-trivial lattices could be made by scatterers with particular symmetry [10 –12]. Although the two monopolar CPA scatterers reported earlier are structurally distinguished [3], their CPA properties are identical. Here we show that there are actually two types of acoustic monopoles. Under single wave excitation, the pressure field near a pressure monopole (PM) at resonance possesses monopolar symmetry, but the velocity field does not have such symmetry, while the velocity field near a velocity monopole (VM) at resonance possesses monopolar symmetry, but the pressure field does not have such symmetry. A hybrid monopole device with neither velocity nor pressure field monopolar symmetry is also demonstrated by combining a VM and a PM, even though it has the mirror structural symmetry. The breaking of monopolar symmetry could lead to higher absorption of the subwavelength scale devices.

The experimental setup for the measurements of the complex transmission coefficient $t$ and the reflection coefficient $r$ under single wave incidence in a 10 cm × 10 cm cross section waveguide is schematically shown in Fig. 1(a-i). Detailed description of the experimental apparatus and procedures can be found elsewhere [8, 13, 14]. Sample-A, which is depicted schematically in Fig. 1(a-ii) and turns out to be a PM resonator, is a hybrid membrane resonator (HMR) shunt on a side wall of the square waveguide [8]. The membrane radius is 25 mm. It is a special version of decorated membrane resonator (DMR) without the decorating platelet [15, 16]. The back cavity is 45 mm in radius and 50 mm in depth. A similar sample was shown in our earlier work as a transparent coherent perfect absorber [3]. Sample-B, which is depicted in Fig. 1(a-iii) and will be shown as a VM resonator with the frequency matching that of Sample-A, is made of two identical DMR's with 22 mm in membrane radius, each mounted on one end of a hollow rigid cylinder 42 mm in length. The central rigid platelet of the DMR's is 60 mg in mass and 3.8 mm in radius. Such structures have been reported earlier as an opaque resonator [3, 14]. The remaining cross section area of the waveguide was sealed off with hard plastic plate when the transmission and reflection spectra of the sample were measured alone under one-side incidence. The sample has a number of dipolar and monopolar resonances. CPA was also observed for a similar sample under the same condition as the PM at its monopolar resonance [3]. Therefore, CPA effect alone cannot fully distinguish the intrinsic symmetry properties of the two monopolar resonators as represented by Sample-A and Sample-B.

Numerical simulations were conducted by COMSOL Multiphysics package using the following parameters. The mass density, Poisson's ratio, pre-stress, and Young's modulus of the membrane were 980 kg/m$^3$, 0.49, 0.5 MPa, and 0.5 MPa. The platelets were much more rigid than the membrane. The speed and the mass density of air used were 343 m/s and 1.29 kg/m$^3$. The thickness of the elastic membranes was 0.02 mm.

The transmission amplitude |t| and the reflection amplitude |r| spectra of Sample-A are shown in Fig. 1(b). Against the background of nearly 100 % transmission and 0 reflection, the spectra are characterized by a series of resonances, each generating a nearly zero impedance boundary across the waveguide cross section and significantly reducing the transmission, accompanied by reflection peaks rising from the near-zero background, as marked by the arrows in the figure at 410.6, 542.7, 661.5, 890.2, and 1056.1 Hz. This is the reason that the sample is called transparent resonator, as the wave in the waveguide is blocked only when the resonator is at resonance [3]. Figure 1(c) depicts the complex scattering parameters $S_+ \equiv r - t$ (red curves) and $S_- \equiv r + t$ (green curves). To be noted is that the complex $S_+$ (solid and dashed red curves) is nearly equal to $-1$, even at the resonant frequencies where $r$ and $t$ undergo large variations, while the $S_-$ exhibits a pronounced resonant feature at each of these frequencies.

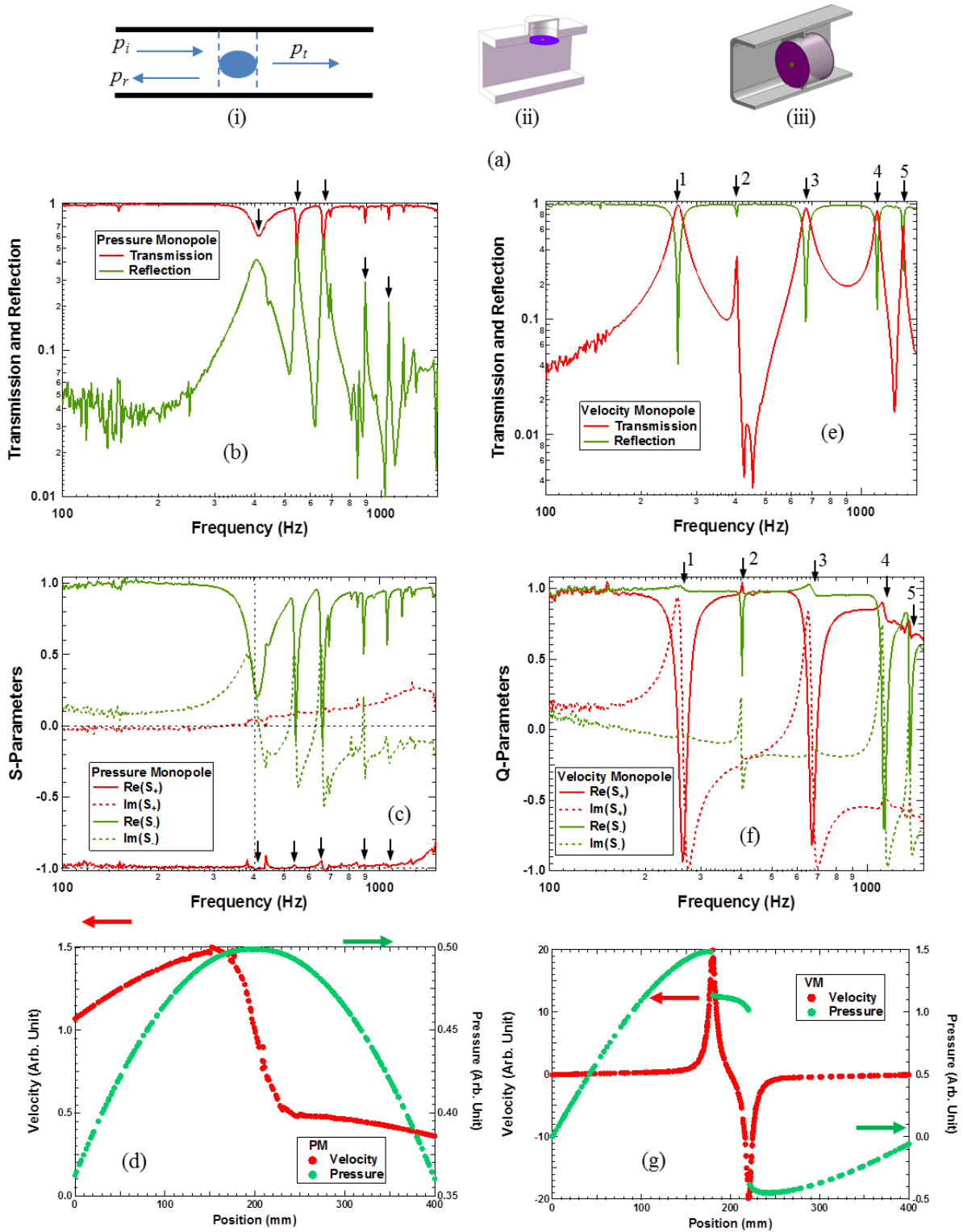

Figure 1 (a-i) Wave scattering scheme by a scatterer under one wave incidence. (a-ii) Schematic structure of Sample-A. (a-iii) Schematic structure of Sample-B. (b) The transmission (red curve) and the reflection (green curve) spectra of Sample-A. The solid arrows mark the resonances of the sample. (c) The scattering parameters of Sample-A obtained from the complex experimental transmission and reflection spectra. (d) The velocity (red disks) and the pressure (green disks) fields averaged over the cross section of the waveguide as a function of position near the sample at the first resonance at 410.6 Hz. The center of the sample is located at the position of the 200 mm mark. (e) The transmission (red curve) and the reflection (green curve) spectra of Sample-B. The solid arrows mark the resonances of the sample. (f) The scattering parameters of Sample-B obtained from the complex experimental transmission and reflection spectra. (g) The velocity (red disks) and the pressure (green disks) fields averaged over the cross section of the waveguide as a function of position near the sample at the first resonance at 403.3 Hz. The center of the sample is located at the position of the 200 mm mark, while that of the two DMR's are at the 180 mm and 220 mm marks, respectively.

The fact that $r - t \equiv S_+ = -1$ at the 410.6 Hz resonance immediately leads to $1 + r = t$, which implies that the total pressure field on the incidence side (1 + r) is equal to that on the transmission side $t$ near the resonator. In other words, the pressure field near the resonator possesses monopolar symmetry even under one-side incidence. The velocity field on the incidence side is proportional to $(1-r)$, and that on the transmission side is proportional to $t$. Obviously, the two equations $1-r=t$ and $1+r=t$ cannot hold simultaneously, unless it is under the trivial off-resonance condition of $r = 0$ and $t = 1$. Therefore, at the 410.6 Hz monopole resonance the velocity field does not possess monopolar symmetry. Indeed, numerical simulation results show that at the resonance, the pressure field averaged over the cross section of the waveguide as a function of position (green points, right axis) has monopolar symmetry relative to the position at 200 mm, which is the center of the HMR, as shown in Fig. 1(d). The velocity field, on the other hand, does not possess monopolar symmetry. For the other four resonances at higher frequencies marked by the arrows, the imaginary part of $S_+$ rises from near zero, indicating small deviation from the monopolar symmetry of the pressure field as characterized by the constraint $S_+ = -1$, most likely due to the small non-symmetric motion [15] of the membrane due to sample imperfection. However, the overall deviation is rather small for all the observed resonances below 1500 Hz. Therefore, at all five resonances the pressure field near Sample-A possesses near monopolar symmetry. Therefore, Sample-A is a *pressure monopole* resonator at 410.6 Hz, and nearly a PM at the other resonant frequencies.

The transmission and the reflection amplitude spectra of Sample-B are shown in Fig. 1(e). Away from the resonant frequencies the sample acted like a hard wall, with nearly total reflection and zero transmission. This is why it is called an opaque resonator. Five resonances are labeled as resonance-1 through 5 as characterized by transmission maxima and reflection minima. The complex scattering parameters $S_+ \equiv r-t$ and $S_- \equiv r+t$ are shown in Fig. 1(f) with the same color and line scheme as for Sample-A in Fig. 1(c). For resonance-1 at 261.9 Hz and resonance-3 at 667.5 Hz, $S_+$ exhibits pronounced resonant features. However, $S_- \equiv r+t$ is nearly equal to 1, which immediately leads to $1-r=t$, implying that the velocity fields near the sample possesses dipolar symmetry. Therefore, we refer to these two resonances as velocity dipolar (VD) resonances. Such VD resonances are common characteristics of single DMR [16], as dictated by their geometric constraint [5]. Here VD resonances appeared in coupled double DMR's also.

For resonance-2 at 403.3 Hz the $S_-$ curves exhibits a strong resonant feature similar in line shape as the PM, while the constraint $S_+ = 1$ is well preserved. This immediately leads to $1-r=-t$, implying that the velocity field possesses monopolar symmetry at resonance-2. Indeed, numerical simulation results in Fig. 1(g) confirm the monopolar symmetry of the velocity field (red points). The pressure field does not possess monopolar symmetry, and exhibits discontinuity at the two DMR's, as expected. Therefore, we refer to resonance-2 as a *velocity monopole* (VM) resonance.

Summarizing the above results, we conclude that there are two types of monopoles, namely PM and VM, which are distinguished by the constraints on their S-parameters. For PM constraint, $S_+^{PM} = -1$, while for VM constraint, $S_+^{VM} = 1$. For VD constraint, $S_-^{VD} = 1$. In the meantime, for monopolar resonances $S_-$ exhibits large resonance features, while for dipolar resonances $S_+$ exhibits large resonance features. According to these constraints and resonant characteristics, we can identify resonance-4 and -5 in Fig. 1(e) as monopolar resonances riding on the strong background of dipolar resonance-3.

We now analyze the underlying mechanism that leads to the constraint conditions. The surface responses of an elastic body, *i. e.*, its displacement at $\vec{r}$ on the surface under a surface excitation at $\vec{r}'$, can be described by the following expression using the Green's function formulism [17]:

$$G(\vec{r},\vec{r}') = \sum_n \frac{u_n(\vec{r})u_n^*(\vec{r}')}{\rho_n(\omega_n^2 - \omega^2 - i\omega\beta_n)} \qquad (1),$$

where $u_n(\vec{r})$ is the vibration field of the n-th eigenmode, $\omega_n$ is the n-th eigenmode angular frequency, $\beta_n$ is the dissipation, $\rho_n \equiv \int \rho(\vec{r}) u_n^2(\vec{r}) dv$, and $\rho(\vec{r})$ is the mass density distribution. For a front-and-back symmetric quasi one-dimensional elastic body like Sample-A and Sample-B occupying the space within $(-a, a)$, the symmetric and the anti-symmetric surface mode functions are given by $u_\pm(x) = \delta(x+a) \pm \delta(x-a)$. For a dipolar response $u_+(x)$ under a dipolar excitation $u_+(x')$, the surface-averaged response function is

$$G_\rho \equiv \iint G(x, x') u_+(x) u_+(x') dx dx' = \sum_n \frac{(u_n(a) + u_n(-a))^2}{2\rho_n(\omega_n^2 - \omega^2 - i\omega\beta_n)} \quad (2a),$$

where $u_n(a)$ is the surface average of $u_n(\vec{r})$ over the cross section at $x = a$. The response function for a monopolar response $u_-(x)$ under a monopolar excitation $u_-(x')$ is

$$G_\lambda \equiv \iint G(x, x') u_-(x) u_-(x') dx dx' = \sum_n \frac{(u_n(a) - u_n(-a))^2}{2\rho_n(\omega_n^2 - \omega^2 - i\omega\beta_n)} \quad (2b).$$

The cross response functions are

$$G_\pm = \iint G(x, x') u_+(x) u_-(x') dx dx' = \sum_n \frac{u_n(a)^2 - u_n(-a)^2}{2\rho_n(\omega_n^2 - \omega^2 - i\omega\beta_n)} \quad (2c),$$

which are usually zero if the elastic body possesses front-and-back mirror structural symmetry.

According to the scattering theory [18], the transmission and the reflection coefficients of an elastic body in a one dimensional waveguide are $S_\pm = r \mp t$. Here $S_+$ and $S_-$ are the symmetric and the anti-symmetric scattering coefficients given by [17] $S_+ = \frac{Z_0 + Z_\rho}{Z_\rho - Z_0} e^{-2ik_0 a}$, $S_- = \frac{Z_\lambda - Z_0}{Z_0 + Z_\lambda} e^{-2ik_0 a}$, where $Z_\rho \equiv \frac{1}{i\omega x_0 G_\rho}$, $Z_\lambda \equiv \frac{1}{i\omega x_0 G_\lambda}$, $k_0$ is the wavevector of sound in air, and $Z_0$ is the acoustic impedance of air. For our samples, $e^{-2ik_0 a} \approx 1$.

For a PM, assuming that the pressure field deviates slightly from perfect monopolar symmetry, so that $S_+ = \frac{Z_0 + Z_\rho}{Z_\rho - Z_0} = -1 + \varepsilon$, ($\varepsilon \ll 1$), then $|G_\rho| = |\frac{2i}{\omega x_0 \varepsilon Z_0}| \gg \frac{1}{Z_0}$. Therefore, the dipolar surface response function of a PM is very large.

As we have seen in Fig. 1(d), the velocity (and the displacement) field of a PM does not possess monopolar symmetry. Therefore, for the dipolar response function in Eq. 2(a) the numerator $u_n(a)+u_n(-a) \neq 0$, while the denominator becomes very small when $\omega_n = \omega$, leading to a very large $G_\rho$. In the meantime, the monopolar response function $G_\lambda$ undergoes expected large variation near the monopole resonance frequency, giving rise to the large resonant feature in $S_-$.

For a VM, assuming that the velocity field near the resonator deviates slightly from perfect monopolar symmetry, i. e., $\frac{Z_0 + Z_\rho}{Z_\rho - Z_0} = 1 + \varepsilon$. Then $|G_\rho| = |\frac{2\varepsilon}{i\omega x_0 Z_0}| \ll |G_{air}|$. This is due to the fact that $u_n(x_0) = -u_n(-x_0)$ for a perfect VM, leading to $G_\rho = 0$. In the meantime, the monopolar response function $G_\lambda$ gives rise to the large resonant feature in $S_-$. Against initial intuition, it is the dipolar response $G_\rho$, rather than $G_\lambda$, of the PM's and VM's that leads to the difference of their symmetry property. For a PM, $G_\rho$ is very large, while for a VM, $G_\rho$ is very small.

For a VD, $\frac{Z_\lambda - Z_0}{Z_0 + Z_\lambda} = S_- = 1$ leads to $Z_\lambda \gg Z_0$ and very small $G_\lambda$, which agrees with known results [16, 17]. An obvious example is a DMR, where the membrane monopolar resonance is at a very high frequency because of the thin membrane, leading to very small $G_\lambda$ near the low frequency dipolar resonances.

To further explore the acoustic symmetry of resonators with mirror symmetric structure we placed one PM (Sample-A) shunt on the waveguide wall and one VM (Sample-B) in the waveguide to form a hybrid monopole (HM), as shown in the insert of Fig. 2(a) as Sample-C. Unlike in the case when Sample-B alone was measured, some cross section area of the waveguide was left open so the shunt HMR of Sample-A could function properly as a PM. Under one wave incidence neither the pressure nor the velocity fields would be symmetric at the original monopole resonance frequency. This is indeed confirmed by experimental results shown in Fig. 2. The transmission and the reflection spectra of the HM shown in Fig. 2(a) exhibit a series of resonances. Resonance-1 at 257.2 Hz, which is close to resonance-1 at 261.9 Hz of Sample-B, is a transparent VD [3], according to the S-parameters in Fig. 2(b) showing that the dipolar constraint $S_+ = 1$ is well preserved, while $S_-$ exhibits a large resonant feature. It is originated from the opaque VD of Sample-B that generates a near-zero impedance boundary across the waveguide, so rather than having maximum transmission and minimum reflection as in the Sample-B case, the transmission displays a minimum and the reflection a maximum.

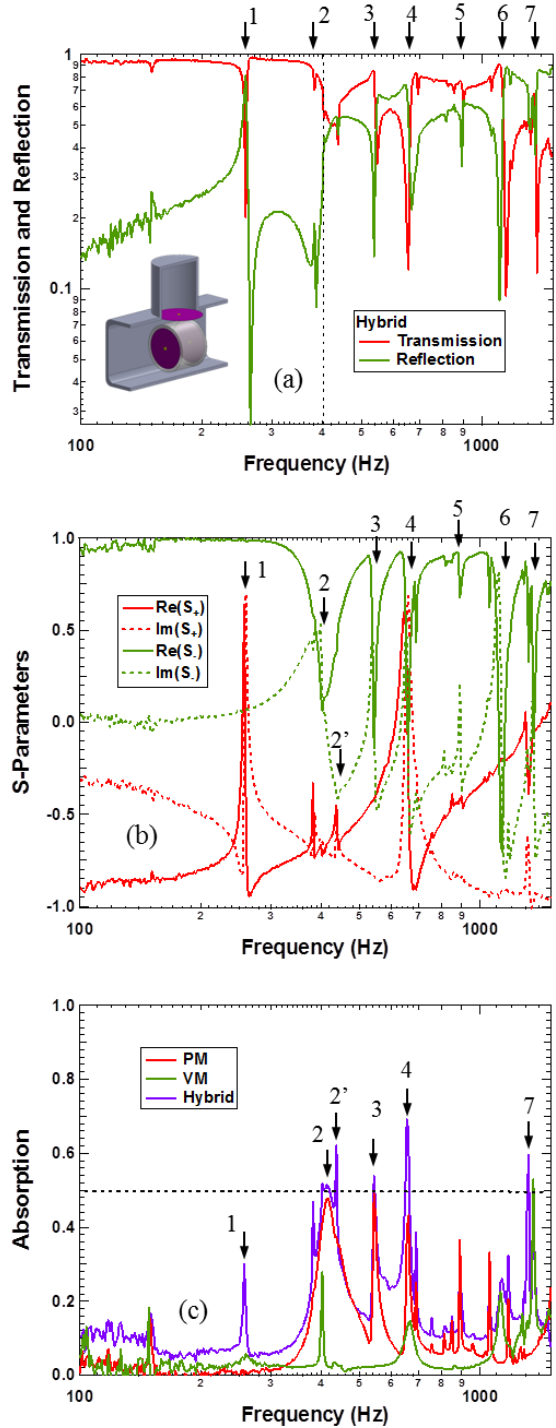

Figure 2    (a) The transmission (red curve) and the reflection (green curve) spectra of Sample-C. The solid arrows mark the resonances of the sample. The insert depicts the schematic structure of the sample. (b) The scattering parameters of Sample-C obtained from the complex experimental transmission and reflection spectra. (c) The absorption spectra of Sample-A (red curve), Sample-B (green curve), and Sample-C (purple curve).

Resonance-2 and resonance-2' combined are a hybrid monopole (HM) resonance of mixed PM and VM, as indicated by the large resonance feature in $S_-$. Instead of featureless, however, the $S_+$ spectrum also exhibits two features riding on the large background of resonance-1 and resonance-4, indicating that even without the influence of the nearby resonance-1 and resonance-4, neither the velocity and nor the pressure fields at this resonance possess monopolar symmetry, i. e, it is a HM resonance. Resonance-3 is a PM with featureless $S_-$ originated from resonance-2 of the HMR in Sample-A, as Sample-B does not have monopole resonance near the frequency. From the strength of the resonance features in both $S_+$ and $S_-$ of resonance-4 one can deduct that it is a degenerated resonance of PM and VD, similar to the one reported earlier [19]. The PM came from the HMR and the VD came from the double membrane Sample-B. Resonance-5 and resonance-6 are PM originated from resonance-4 and resonance-5 of Sample-A. Resonance-7 is another hybrid PM and VM resonance riding on the large background of resonance-4. The VM is originated from resonance-5 of Sample-B, and the PM is probably originated from a weak PM resonance beyond resonance-5 of Sample-A, enhanced due to the reduced clear cross section portion of the waveguide.

It is straightforward to show that the symmetry constraints limit the maximum absorption of the resonators to 0.5 [5, 6]. Breaking of such symmetry constraints could lead to higher absorption [7 – 9, 19]. Shown in Fig. 2(c) are the experimental absorption spectra of Sample-A (red curve), Sample-B (green curve), and Sample-C (purple curve). The absorption peak near 500 Hz of Sample-A is slightly above the limit value of 0.5, probably due to the small asymmetric motion of the membrane caused by uneven tension in the membrane, while the $S^+ = -1$ constraint is slightly broken. Two absorption peaks of Sample-C exceed the 0.5 limit by an obvious margin. The absorption of the HM resonance at 437 Hz reaches 0.6. The degenerate monopole-dipolar resonance at 655.6 Hz even reached 0.69, showing clearly the effect of enhanced absorption by breaking the monopolar and dipolar symmetry constraints. The absorption peak of the degenerate monopole-dipolar in the present work is not as high as that in Ref. 19, which reached the near perfect value of 1, because here the relative strength of the PM and the VD resonances are not intentionally optimized as in Ref. 19.

The concept of PM and VM symmetry, and HM for the breaking of PM and VM symmetry in quasi one dimensional system reported here could be extended to two dimensional counterparts without much modifications in the structural designs. For example, a HMR shunt on the top or the bottom plate in a two dimensional waveguide could serve as a PM, and enclosed membrane structures with decorating platelets could serve as VM and VD. Quadruples and higher order multipoles with pressure or velocity symmetry could also be constructed with proper combinations of simpler units. The acquiring or breaking of the symmetry of the velocity field and/or the pressure field could add additional options to the designs of systems with specific topological symmetry, doubly-negative effective parameters, and other novel properties. The potential of exploring the symmetry properties of vibration fields near elastic structures could be very rewarding.

**Acknowledgement**

This work was supported by AoE/P-02/12 from the Research Grant Council of the Hong Kong SAR government.